%% file: paper.tex
\documentclass{emulateapj}
\submitted{Draft version  \today}
\journalinfo{\@submitted}
\usepackage{natbib}
\usepackage{xcolor}
\usepackage{color}
\usepackage{comment}
\definecolor{color}{RGB}{25,25,112}
\definecolor{negro}{RGB}{0,0,0}
\definecolor{colorurl}{RGB}{25,25,112}

\usepackage[hyperfootnotes=false]{hyperref}
\hypersetup{
  colorlinks,
  citecolor=color,
  linkcolor=negro,
  urlcolor=colorurl}

 \newcommand{\be}{\begin{equation}}
\newcommand{\ee}{\end{equation}}
\newcommand{\bse}{\begin{subequations}}
\newcommand{\ese}{\end{subequations}}
\newcommand{\bary}{\begin{eqnarray}}
\newcommand{\eary}{\end{eqnarray}}

\shorttitle{Photometric observations of Supernova  2013cq}
\shortauthors{Becerra et al.}

\begin{document}

\title{Photometric Observations of Supernova  2013cq Associated with GRB 130427A}
 \author{%
Becerra, R. L.$^{1}$,
Watson, A. M.$^{1}$,
Lee, W. H.$^{1}$;
Fraija, N.$^{1}$;
Butler, N. R.$^{2}$;
Bloom, J. S.$^{3}$; 
Capone, J. I.$^{4}$; 
Cucchiara, A.$^{5}$; 
de Diego, J. A.$^{1}$; 
Fox, O. D.$^{6}$
Gehrels, N.$^{7}$;  
Georgiev, L. N.$^{1}$;
Gonz\'alez, J. J.$^{1}$;  
Kutyrev, A. S.$^{7}$; 
Littlejohns, O. M.$^{2}$;
Prochaska, J. X.$^{8}$; 
Ramirez-Ruiz, E.$^{8}$; 
Richer, M. G.$^{9}$;  
Rom\'an-Z\'u\~niga, C. G.$^{9}$; 
Toy, V. L.$^{4}$; 
Troja, E.$^{4,7}$ 
}
\address{$^1$ Instituto de Astronom\'ia, Universidad Nacional Aut\'onoma de M\'exico, Apartado Postal 70-264, 04510 M\'exico, D. F., M\'exico;\\
$^2$ School of Earth and Space Exploration, Arizona State University, Tempe, AZ 85287, USA;\\
$^3$Department of Astronomy, University of California, Berkeley, CA 94720-3411, USA;\\
$^4$Department of Astronomy, University of Maryland, College Park, MD 20742, USA;	\\
$^5$NASA Postdoctoral Program Fellow, Goddard Space Flight Center, Greenbelt, MD 20771, USA;\\
$^6$Space Telescope Science Institute, 3700 San Martin Drive, Baltimore, MD 21218, USA;\\
$^7$NASA, Goddard Space Flight Center, Greenbelt, MD 20771, USA;\\
$^8$Department of Astronomy and Astrophysics, UCO/Lick Observatory, University of California, 1156 High Street, Santa Cruz, CA 95064, USA;\\
$^9$Instituto de Astronom\'ia, Universidad Nacional Aut\'onoma de M\'exico, Apartado Postal 106, 22800 Ensenada, Baja California, M\'exico;\\
}

\begin{abstract}
We observed the afterglow of GRB 130427A with the RATIR instrument on the 1.5-m Harold L. Johnson telescope of the Observatorio Astron\'omico Nacional on Sierra San Pedro M\'artir. Our homogenous $griZYJH$ photometry extends from the night of burst to three years later. We fit a model for the afterglow.  There is a significant positive residual which matches the behavior of SN 1998bw in the $griZ$ filters; we suggest that this is a photometric signature of the supernova SN 2013cq associated with the GRB. The peak absolute magnitude of the supernova is $M_r=-18.43\pm0.11$.\\[2ex]
\end{abstract}
\begin{center}
\end{center}

\section{Introduction}

Gamma-ray bursts (GRBs) are the most energetic events in the universe and are  produced at cosmological distances. They can be classified according to their duration $T_{90}$, the time interval in the observer's frame over which 90\% of the total background-subtracted counts are observed \citep{1995AAS...186.5301K}. This parameter has long pointed to a bi-modal distribution \citep{Kouveliotou93}.

Long GRBs ($T_{90}> 2$ s) are today thought to be the result of the core-collapse of a star \citep{1993ApJ...3505..273W,mw99,2012grbu.book..169H} with an initial mass with more than 10 M$_{\odot}$   \citep[see][for a review]{wb06}, while short GRBs ($T_{90} < 2$ s) are thought to be the result of mergers between two compact objects \citep{ls76,bp86,bp91,eichler89,npp92} like black holes or neutron stars \citep[see][for reviews]{2007NJPh....9...17L,nakar07}. 

\cite{1993ApJ...3505..273W} specifically proposed a way in which the core collapse of massive stars could lead to a long GRB, and thus be possibly associated with a supernova (SN) \citep{2011AN....332..434M,2006ARA&A..44..507W,2012arXiv1206.6979B,2013RSPTA.37120275H}. In this scenario, the optical emission from the SN would appear a few days after the GRB, when the ejecta becomes optically thin. This leads to two ways to identify the presence of a SN associated with a GRB. First, by the appearance of the broad spectral lines characteristic of SN a few days after the burst, and second, through a rebrightening in the lightcurve of the GRB after a few days due to the broad-band emission of the SN. The identification of a SN associated with GRB 980425  \citep{1998Natur.395..670G} showed that at least some GRBs are truly linked to the core collapse of massive stars. 
   
Subsequently, other SNe have also been associated with long GRBs. Confirmed spectroscopic cases are listed in Table \ref{tabla:grb-sn}. Most or perhaps all of these SNe are type Ic. Usually the hosts of GRB-SNe are blue, star-forming galaxies \citep{2000ApJ...542L..89F,2006AA...447..891F,2006AA...454..103H,Fruchter06,2013EAS....61..427N} and the events occur within a low metallicity environment \citep{2008AJ....135.1136M}.

Simply taking into account the rates of SNe and GRBs, however, it is apparent that not all core-collapse SNe produce long GRBs, and special conditions are required in order to successfully power a burst. These probably involve rotation, magnetic fields, chemical composition, binarity, or a combination of the above and are not yet fully resolved. Thus, precise observations of a significant sample of GRBs associated with SNe are fundamental in order to determine the evolutionary pathways that can lead to such a link. Given that  the current sample is extremely limited, studying single events like SN 2013cq associated with GRB 130427A in great detail offers the opportunity to provide unique additional insights and to eventually lead the way to a statistically significant sample, and this is the main motivation for the present paper.

GRB 130427A is one of the most brightest gamma-ray burst in the last few years, and had $E_{\gamma,iso}=1.40\times 10^{54}$~erg in total isotropic energy release \citep{2014Sci...343...352A} and $E_{\rm \gamma,peak}$ = $1028 \pm 8$ keV \citep{2014Sci...343...358M}. It was detected at high energies by several satellite-borne instruments and lead to a flurry of ground-based observations. In total, there have been 91 GCN Circulars related to GRB 130427A. Its redshift was measured to be $z=0.34$ \citep{2013GCN..14455...1L}. RAPTOR (Rapid Telescope for Optical Response) observed a bright optical flash with a magnitude of $7.03\pm 0.03$ in the time interval from 9.31 s to 19.31 s after the GBM trigger \citep{2014Sci...343...38V}.  The bright optical flash at early times was modeled with synchrotron emission from reverse shocks \citep{2014Sci...343...38V, 2016ApJ...818..190F}. \cite{2014ApJ...781...37P} show multi-wavelength optical/infrared photometry of the afterglow of GRB 130427A, and explain the afterglow through synchrotron radiation and suggest a massive-star progenitor. Spectroscopy with the 10.4-m GTC telescope reported by \cite{2013ApJ...776...98X} showed a broad-lined Ic SN 2013cq associated with GRB 130427A.

 This paper presents a detailed set of calibrated and uniform photometry of the bright GRB 130427A with the RATIR instrument in the $griZYJH$ filters. The major advantages of our work compared to earlier papers \citep{2014ApJ...781...37P,2013ApJ...776...98X,2014A&A...567A..29M} is that our photometry is generally deeper, has better temporal sampling, and we subtract the host galaxy using deep late-epoch image. Furthermore, our data were all obtained with the same instrument, using the same observing strategy, and were all processed in the same way. This means that our data are naturally homogeneous. The paper is organized as follows: in \S\ref{sec:observations} we present the observations; in \S\ref{sec:model} we fit the data using segments of power-law according to the fireball model;

 \emph{in \S\ref{sec:results} search for the signature of SN 2013cq in the the difference between the host-subtracted measurements and the power-law afterglow model;}

in \S\ref{sec:discussion} we discuss the results and summarize our conclusions.

\section{Observations}
\label{sec:observations}

\subsection{Fermi and Swift}

The Gamma-Ray Burst Monitor (GBM) instrument on the Fermi satellite triggered on GRB 130427A at 07:47:06.42 UTC on 2013 April 27 \citep{2013GCN..14473...1V}. Subsequently, the Burst Alert Telescope (BAT) on the Swift satellite triggered on the GRB at 07:47:57.51 UTC \citep{2014Sci...343...358M}. The duration measured with BAT was $T_{90}=163$ s \citep{2013GCN..14470...1B}, making GRB 130427A a long GRB. 

\subsection{RATIR}

The Reionization and Transients Infrared Camera (RATIR) is a four-channel simultaneous optical and near infrared imager mounted on 1.5-m Harold L. Johnson Telescope at the Observatorio Astron\'omico Nacional on Sierra San Pedro M\'artir in Baja California, Mexico. RATIR responds autonomously to GRB triggers from the Swift satellite and obtains simultaneous photometry in $riZJ$ or $riYH$ \citep{2012SPIE.8446E..10B,2012SPIE.8444E..5LW,2015MNRAS.449.2919L}. In manually-programmed follow-up observations, the $g$ filter can be substituted for $r$.

RATIR began to observe the field of GRB 130427A 15.5 minutes after the BAT trigger, and continued to observe it intensively over the subsequent weeks.  On the first night, the $r$ detector failed, so we only have data in $iZYJH$. On subsequent nights, we have data in $riZYJH$. After one week, we began to observe in $g$ as well. We reobserved the field on several nights in 2014 and 2016 mainly to place constraints on the host galaxy. 

Our reduction pipeline performs bias subtraction and flat-field correction, followed by astrometric calibration using the astrometry.net software \citep{2010AJ....139.1782L}, iterative sky-subtraction, coaddition using SWARP, and source detection using SEXTRACTOR \citep{2015MNRAS.449.2919L}. We calibrate against SDSS and 2MASS \citep{2015MNRAS.449.2919L}. The systematic calibration error is about 1\%.

The individual exposures were 80 s in $gri$ and 67 s in $ZYJH$ filters (with the infrared exposures being shorter because of their longer read-out overhead). On the first night, we consider the exposures individually. For the second to the fifth night, we combined sets of 16 exposures taken over about 30 minutes to improve the signal-to-noise ratio. For the remaining nights in 2013, we combined all of the exposures for each night, for 2014, we combined several nights, and for 2016, we combined all of the exposures. The image quality in the final images was typically 2 arcsec FWHM.

We obtained aperture photometry using a 3 arcsec diameter aperture. Table~\ref{tab:datos} gives our aperture photometry. For each image it gives the start and end time, $t_0$ and $t_f$, the total exposure time $t_e$, the magnitude, the 1$\sigma$ total uncertainty (including both statistical and systematic contributions), and the filter. These magnitudes are not corrected for Galactic extinction.

\begin{figure*}
\centering
 \includegraphics[width=1\textwidth]{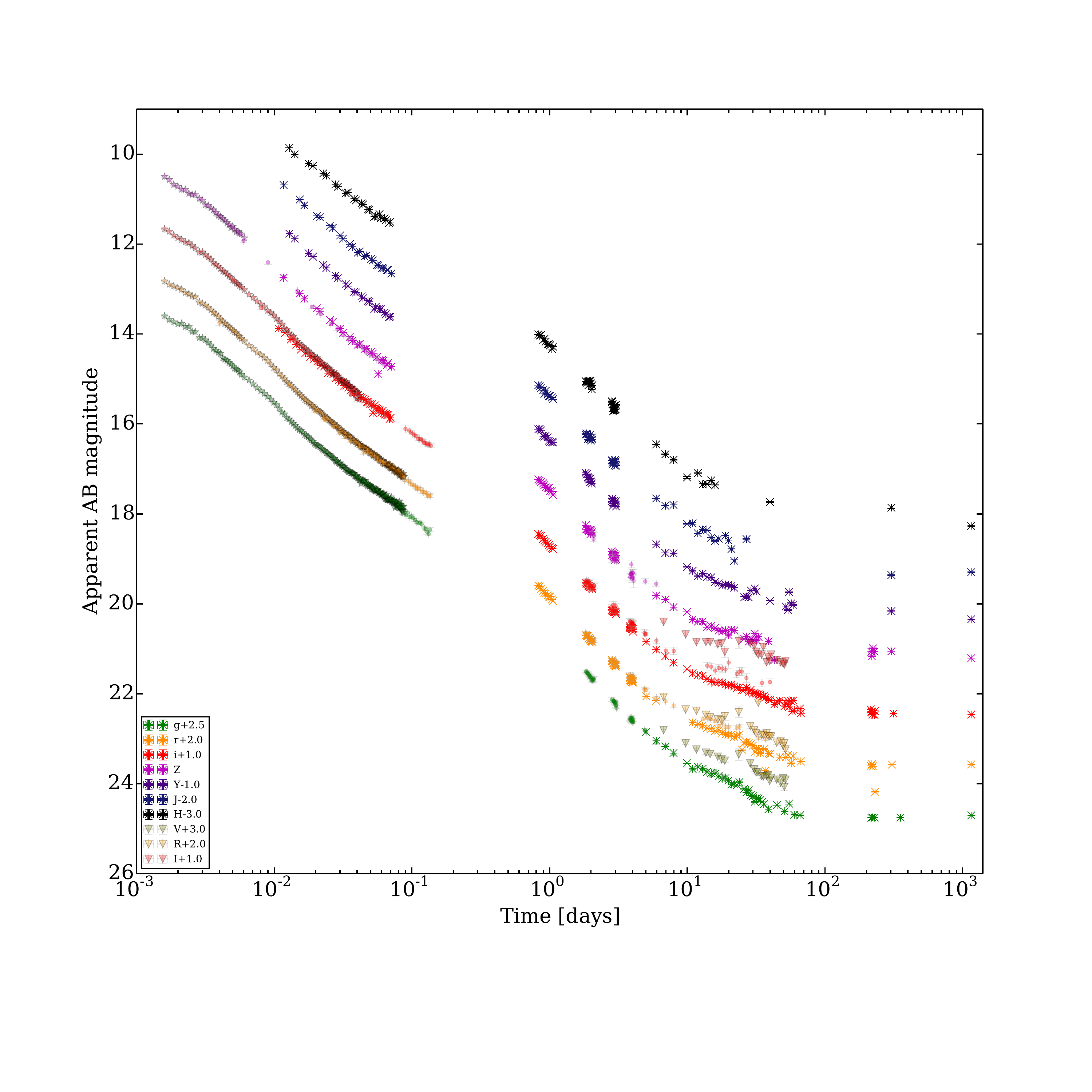}
 \caption{ $griZYJH$ light curves for GRB 130427A from RATIR (lines), RAPTOR \citep{2014Sci...343...358M} (stars) and Palomar P60  (points) \citep{2014ApJ...781...37P} and VLT  \citep{2014A&A...567A..29M} (triangles). The RATIR photometry shown here is our aperture photometry of the afterglow and host galaxy.}
 \label{fig:flujos}
\end{figure*}

We also obtained PSF-fitting photometry of the afterglow and supernova in our 2013 and 2014 images after subtracting the host galaxy using our 2016 image. For each image, we estimated the point-spread function (PSF) by co-aligning and summing images of stars (as categorized by the SDSS) within 3 arcmin of the GRB. We subtracted the 2016 image from the earlier images using HOTPANTS \citep{2015ascl.soft04004B} and fitted the PSF to the residual. Even though our image quality is typically 2 arcsec FWHM, we cannot reliably perform PSF-fitting on the unsubtracted images because the galaxy is offset about 0.8 arcsec to the south-east of the afterglow \citep{2014ApJ...792..115L}. Table \ref{tab:datos} also gives our PSF-fitting photometry. The main advantage of PSF-fitting is that the statistical uncertainties are reduced typically by about 20\%.
Figure \ref{fig:flujos} shows the RATIR optical and near-infrared light curves.

\begin{deluxetable*}{llclcl}
\tabletypesize{\scriptsize}
\tablewidth{0pt}
\tablecaption{GRBs with associated SN \label{tabla:grb-sn}}
\tablehead{\colhead{GRB}&\colhead{SN}&\colhead{SN type}&\colhead{z}&\colhead{Evidence\tablenotemark{a}}&\colhead{References}}
 GRB 980425&SN 1998bw&Ic&0.0085&A&1\\
 GRB 011121&SN 2001ke&IIn?/Ic?\tablenotemark{b}&0.362&B&2, 3, 4\\
 GRB 021211&SN 2002lt&Ic&1.006&B&5\\
 GRB 030329&SN 2003dh&Ic&0.1687&A&6, 7\\
 GRB 031203&SN 2003lw&Ic&0.105&A&8, 9, 10\\
 GRB 050525A&SN 2005nc&Ic&0.606&B&11\\
 GRB 060218&SN 2006aj&Ic&0.0335&A&12, 13, 14, 15, 16\\
 GRB 081007&SN 2008hw&Ic&0.530&B&17, 18, 19\\
 GRB 091127&SN 2009nz&Ic&0.49&A&20, 21\\
 GRB 100316D&SN 2010bh&Ic&0.059&A&22, 23, 24, 25, 26\\
 GRB 120422A&SN 2012bz&Ic&0.283&A&27, 28\\
 GRB 130427A&SN 2013cq&Ic&0.34&A&29, 30, 31\\
 GRB 130702A&SN 2013dx&Ic&0.145&A&32, 33\\
 GRB 140606B&iPTF14bfu&Ic&0.384&A&34
 \enddata
\tablenotetext{a}{Evidence for the GRB-SN association, according to the authors, with  A meaning strong spectroscopic evidence and B meaning
a clear light curve bump together with some spectroscopic evidence resembling a
SN.}
\tablenotetext{b}{SN 2001ke has no clear spectroscopic classification. \cite{2003ApJ...582..924G} suggest that it is a type IIn but \cite{2002ApJ...572L..45B} claim it is indeed consistent with a 1998bw-like type Ic.}
\tablerefs{
(1) \cite{1998Natur.395..670G}; (2) \cite{2002ApJ...572L..45B}; (3) \cite{2003ApJ...582..924G}; (4) \cite{2003Natur.426..157G}; (5) \cite{2003AA...406L..33D}; (6) \cite{2003ApJ...591L..17S}; (7) \cite{2003ApJ...599..394M}; (8) \cite{2004ApJ...609L...5M}; (9)\cite{2004ApJ...609L..59G}; (10) \cite{2004AA...419L..21T}; (11) \cite{2006ApJ...642L.103D}; (12) \cite{2006Natur.442.1008C}; (13) \cite{2008AJ....135.1136M}; (14) \cite{2006ApJ...643L..99M}; (15) \cite{2006AA...457..857F}; (16) \cite{2006Natur.442.1011P}; (17) \cite{2008GCN..8335....1B}; (18) \cite{2008CBET.1602....1D};(19) \cite{2008GCN..8662....1S}; (20) \cite{2010ApJ...718L.150C}; (21) \cite{2011ApJ...743..204B}; (22) \cite{2011MNRAS.411.2792S}; (23) \cite{2012ApJ...753...67B}; (24) \cite{2012AA...539A..76O}; (25) \cite{2011ApJ...740...41C}; (26) \cite{2010arXiv1004.2262C}; (27) \cite{2012AA...547A..82M}; (28) \cite{2014AA...566A.102S}; (29) \cite{2013GCN..14646...1D}; (30) \cite{2013ApJ...776...98X}; (31) This work; (32) \cite{2015arXiv150800575T}; (33) \cite{2015AA...577A.116D}; and (34) \cite{2015MNRAS.452.1535C}.}
\end{deluxetable*}

\section{Models}
\label{sec:model}

The standard fireball model for GRBs \citep{2015PhR...561....1K} distinguishes two stages: the \emph{prompt emission} and the \emph{afterglow}. The prompt emission is simultaneous with emission in gamma rays and is produced by internal shocks in the jet driven by the central engine. The afterglow is produced by the external shock between the jet and the circumstellar environment \cite[e.g.][]{2000APJ...532....286K,  2015ApJ...804..105F}. 

The emission region of the radiation determines the form and behavior of the spectrum and light curve for a GRB \cite[e.g.][]{2016ApJ...831..22F}, and can be different for each filter. Optical radiation has three possible origins: internal shocks in the jet, the forward external shock, and the reverse external shock \citep{1999ApJ...520..641S}.

The afterglow phase can be explained by assuming a power-law energy distribution of shocked relativistic electrons, $N(E)\propto E^{-p}$, which lead to the observed flux being a series of power-laws segments as a function of time $t$ and frequency $\nu$ as $F_{\nu}^{\rm syn}\propto t^{-\alpha}\nu^{-\beta}$ \citep{1998ApJ...3597L..17S}.

RATIR began to observe after the end of the prompt emission, so we only have photometry for the afterglow. We divided these data into two epochs: the \emph{early} afterglow for time $t<0.1$ days (the first night) and the \emph{late} afterglow for the $t> 0.7$ day (the second and subsequent nights). This division was guided by the analysis of \cite{2014ApJ...781...37P}, who report a change in the slope of the light curve at $t=0.7$ days. We have no data between $t = 0.1$ and $t = 0.7$ days.

\subsection{Host galaxy}
\label{subs:hostgalaxy}


 \cite{2013GCN..14516...1V} suggested that SDSS DR12 galaxy object 1237667431180861948 was the host galaxy of the GRB. This was subsequently confirmed by the close agreement in redshift between absorption lines in the GRB spectrum \citep{2013GCN..14455...1L,2013GCN..14478...1X,2013GCN..14491...1F} and emission lines from the galaxy \citep{2013ApJ...776...98X}.
 



\cite{2014ApJ...792..115L} obtained \emph{HST} images of the afterglow and host galaxy. They suggested that the host is a moderately star-forming, possibly interacting, disk galaxy and the GRB occured about 0.8 arcsec (4 kpc) from the nucleus.

Table \ref{tab:host} reports our RATIR \emph{griZYJH} aperture magnitudes from 2016 and magnitudes from the SDSS DR12 image using the same apertures and calibrating stars. The magnitudes from our image are consistent with the magnitudes from the SDSS image, but have lower uncertainties.

We can estimate the rest-frame $g-i$ color and $M_i$ magnitude from our observed $r-Y$ color and $Y$ magnitude (see Table~\ref{tabla:bandas} for the correspondence between rest-frame and observed bands) assuming a $\Lambda$CDM model with a $H_0=67.8$ km/Mpc/s \citep{2014A&A...571A...1P}. We obtain a rest-frame $g-i=0.23 \pm 0.08$ and a rest-frame $M_i=-19.91\pm 0.07$. These properties place it among the most extremely blue galaxies in the $z\approx 0$ sample of \cite{2010A&A...517A..73G}.

\begin{deluxetable}{lcccc}
\tabletypesize{\scriptsize}
\tablewidth{0pt}
\tablecaption{Host-galaxy magnitudes for GRB 130427A\label{tab:host}}
\tablehead{\colhead{Filter}&\colhead{$m_\mathrm{RATIR}$}&\colhead{Exposure (h)}&\colhead{$m_\mathrm{SDSS}$}}
\startdata
$g$ & 22.20 $\pm$ 0.04 & 10.6&$22.14\pm0.12$\\
$r$ & 21.57 $\pm$ 0.04  &\phantom{0}6.5&$21.41\pm0.10$\\
$i$ &21.46 $\pm$ 0.03   &17.8&$21.53\pm0.23$\\
$Z$ & 21.29 $\pm$ 0.05 & \phantom{0}8.3&\nodata\\
$Y$ & 21.34 $\pm$ 0.07 &\phantom{0}8.2&\nodata\\
$J$ & 21.30 $\pm$ 0.09  &\phantom{0}8.0&\nodata\\
$H$ & 21.27 $\pm$ 0.13  &\phantom{0}7.9&\nodata\\
\enddata
\end{deluxetable}

\subsection{Early Afterglow ($t<0.7$ days)}
\label{subs:early}

To characterize the early afterglow data ($t<0.7$ days), we used the $iZYJH$ aperture photometry from Table~\ref{tab:datos} as the contribution of the host galaxy can be neglected at early times. We fitted the flux densities with a  power-law model $F=A_\mathrm{E}t^{-\alpha_\mathrm{E}}$, in which $F$ is the flux density in the filter, $A_\mathrm{E}$ is a constant, $t$ is the time since the BAT trigger (in days), and $\alpha_\mathrm{E}$ is the temporal index, assumed to be the same for all filters. The model has six free parameters: the five values of $A_\mathrm{E}$ and the one value of the index $\alpha_\mathrm{E}$.

We minimized the value of $\chi^2$ to find the best-fit parameters. 
The final fit has a $\rm \chi^2/n=1.64$ with $n = 100$ degrees of freedom. The best-fit parameters are given in Table \ref{tabla:fit}. 
The errors were calculated using the standard deviation of the best-fits parameters after Gaussian perturbations around the flux values observed over 10,000 trials. Table \ref{tab:datosres_early} shows the residuals to the fit in the sense of data minus model in units of $\mu$Jy.

\begin{deluxetable}{ccc}
\tabletypesize{\scriptsize}
\tablewidth{0pt}
\tablecaption{Fit parameters\label{tabla:fit}}
\tablehead{\colhead{Parameter}&{Band}&{Value}}
\startdata
 $\alpha_\mathrm{E}$&&$0.97\pm0.04$\\
 $A_\mathrm{E}$&$i$&$303\pm28\mu$Jy\\
 $A_\mathrm{E}$&$Z$&$348\pm38\mu$Jy\\
 $A_\mathrm{E}$&$Y$&$373\pm38\mu$Jy\\
 $A_\mathrm{E}$&$J$&$364\pm37\mu$Jy\\
 $A_\mathrm{E}$&$H$&$373\pm27\mu$Jy\\
 \hline
  $\alpha_\mathrm{L}$&&$1.41\pm0.04$\\
  $A_\mathrm{L}$&$g$&$228\pm12\mu$Jy\\
 $A_\mathrm{L}$&$r$&$268\pm10\mu$Jy\\
 $A_\mathrm{L}$&$i$&$315\pm20\mu$Jy\\
 $A_\mathrm{L}$&$Z$&$388\pm37\mu$Jy\\
 $A_\mathrm{L}$&$Y$&$436\pm18\mu$Jy\\
 $A_\mathrm{L}$&$J$&$426\pm11\mu$Jy\\
 $A_\mathrm{L}$&$H$&$489\pm116\mu$Jy
\enddata
\end{deluxetable}

\subsection{Late Afterglow ($t > 0.7$ days)}
\label{subs:late}

To characterize the late afterglow ($t > 0.7$ days), we use the $griZYHJ$ flux densities from our PSF-fitting photometry of the subtracted images from Table~\ref{tab:datos}. We fitted with a model $F=A_\mathrm{L}t^{-\alpha_\mathrm{L}}$, in which $F$ is the flux density in each filter, $A_\mathrm{L}$ is a constant, $t$ is the time since the BAT trigger (in days) and $\alpha_\mathrm{L}$ is the temporal index, assumed to be the same for all filters. The model has eight free parameters: the seven values of $A_\mathrm{L}$ and the one value of the index $\alpha_\mathrm{L}$.

Again, we minimized the value of $\chi^2$ to find the best-fit parameters. To avoid the worst contamination from the SN, we fitted only the data points from 0.7 to 7 days and from 40 days onwards. The final fit has a $\chi^2/n=1.05$ with $n = 327$ degrees of freedom. The best-fit parameters are given in Table \ref{tabla:fit}. The errors were calculated using the standard deviation of the best-fits parameters with Gaussian perturbations to the flux value and 10,000 trials. Figures \ref{fig:sn_g} to \ref{fig:sn_H} show the data and the best fit. Table \ref{tab:datosres} shows the residuals to the fit in the sense of data minus model in units of $\mu$Jy.

\section{Results}
\label{sec:results}

\subsection{SN Component}
\label{subs:sncomponent}

\begin{figure*}
\centering
 \includegraphics[height=0.30\textheight]{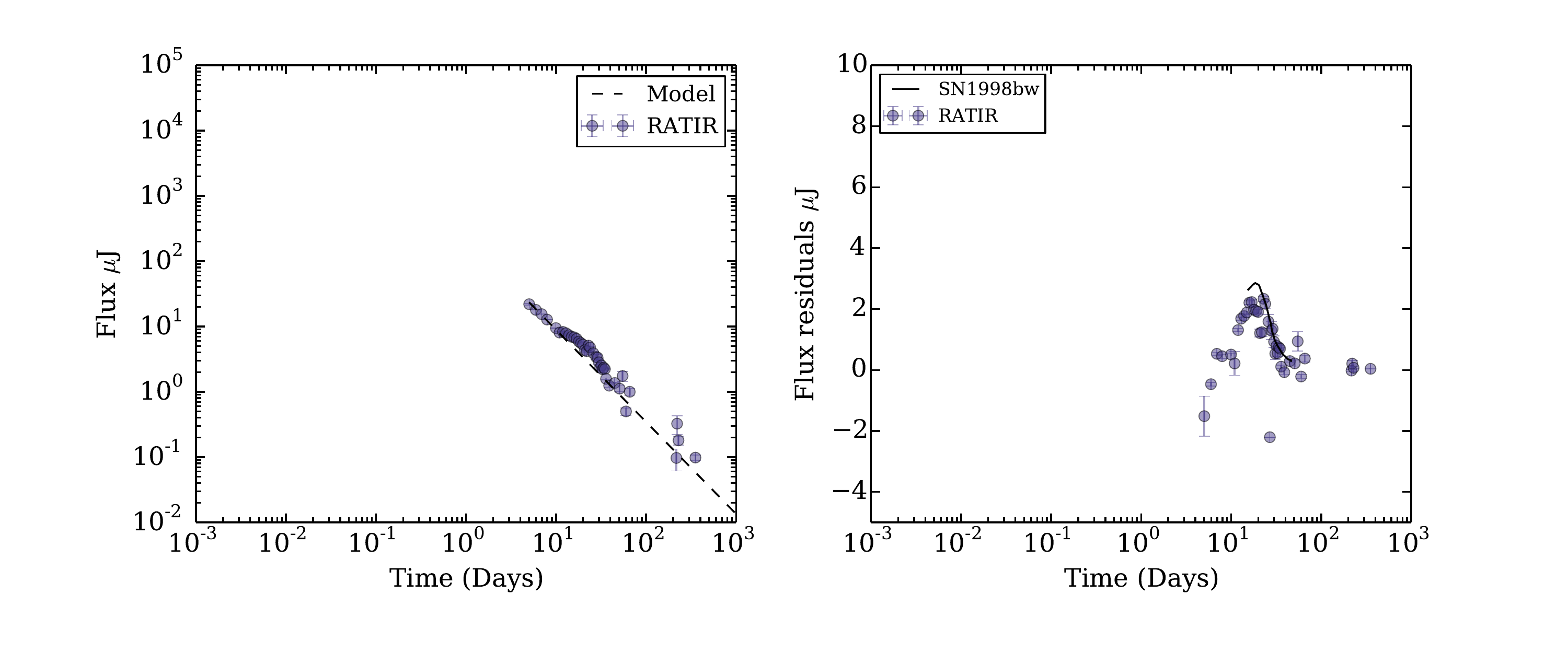}\\[5ex]
 \caption{\emph{Left:} The host-subtracted data (points) and power-law model (dashed line) in $g$ filter of RATIR (points). \emph{Right:} Flux density residuals in $g$ (points) and the flux density of SN 1998bw in $U$ shifted to $z = 0.34$ (continuous line).\\}
 \label{fig:sn_g}
\end{figure*}

\begin{figure*}
\centering
 \includegraphics[height=0.30\textheight]{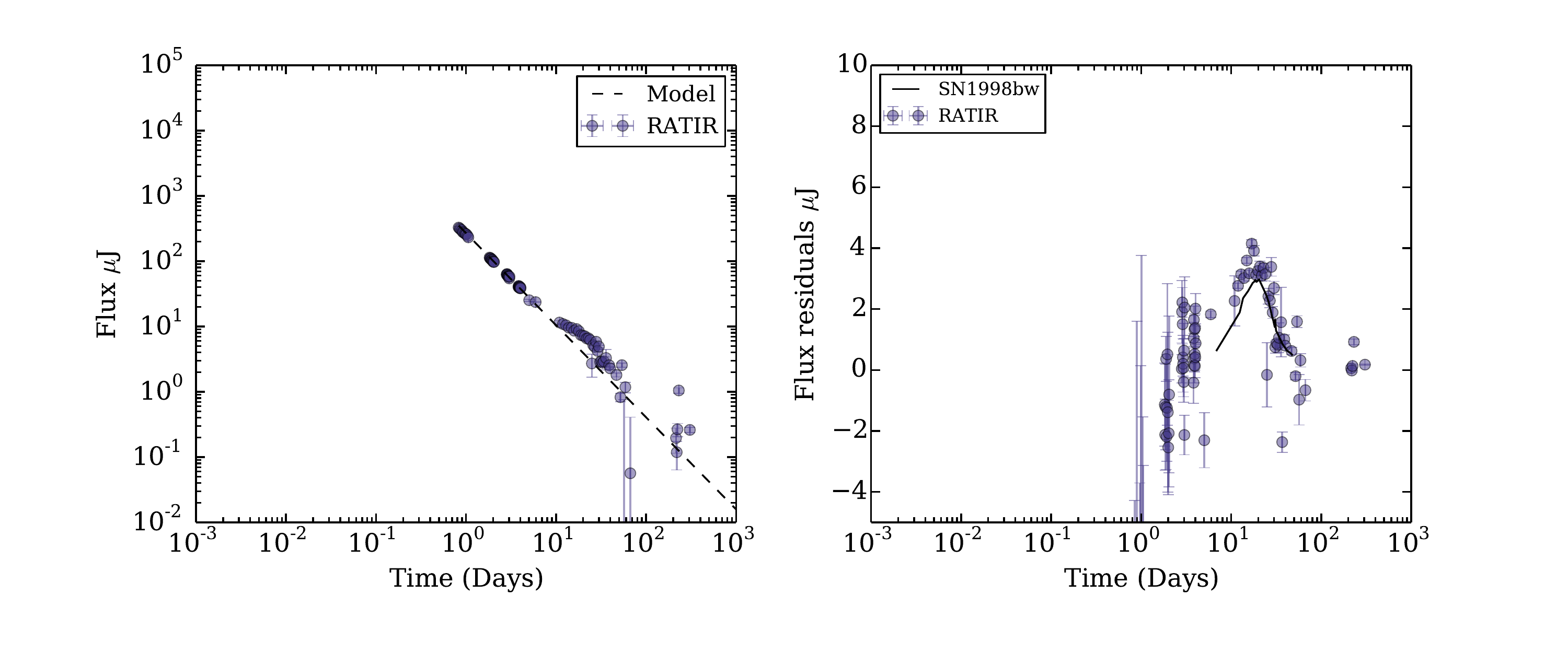}\\[5ex]
 \caption{\emph{Left:} The host-subtracted data (points) and power-law model (dashed line) in $r$ filter of RATIR (points). \emph{Right:} Flux density residuals in $r$ (points) and the flux density of SN 1998bw in $B$ shifted to $z = 0.34$ (continuous line).\\}
   \label{fig:sn_r}
\end{figure*}

\begin{figure*}
\centering
 \includegraphics[height=0.30\textheight]{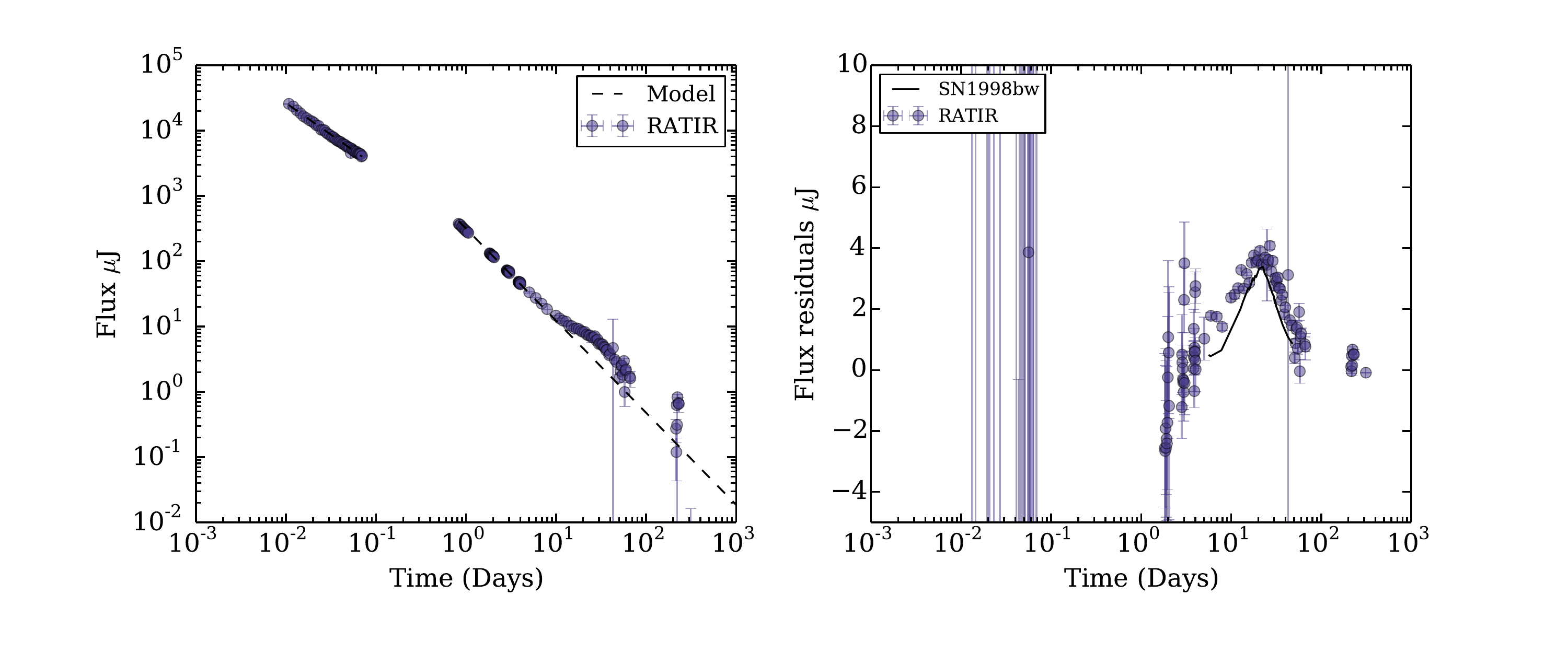}\\[5ex]
 \caption{\emph{Left:} The host-subtracted data (points) and power-law model (dashed line) in $i$ filter of RATIR (points). \emph{Right:} Flux density residuals in $i$ (points) and the flux density of SN 1998bw in $V$ shifted to $z = 0.34$ (continuous line).\\}
   \label{fig:sn_i}
\end{figure*}

\begin{figure*}
\centering
 \includegraphics[height=0.30\textheight]{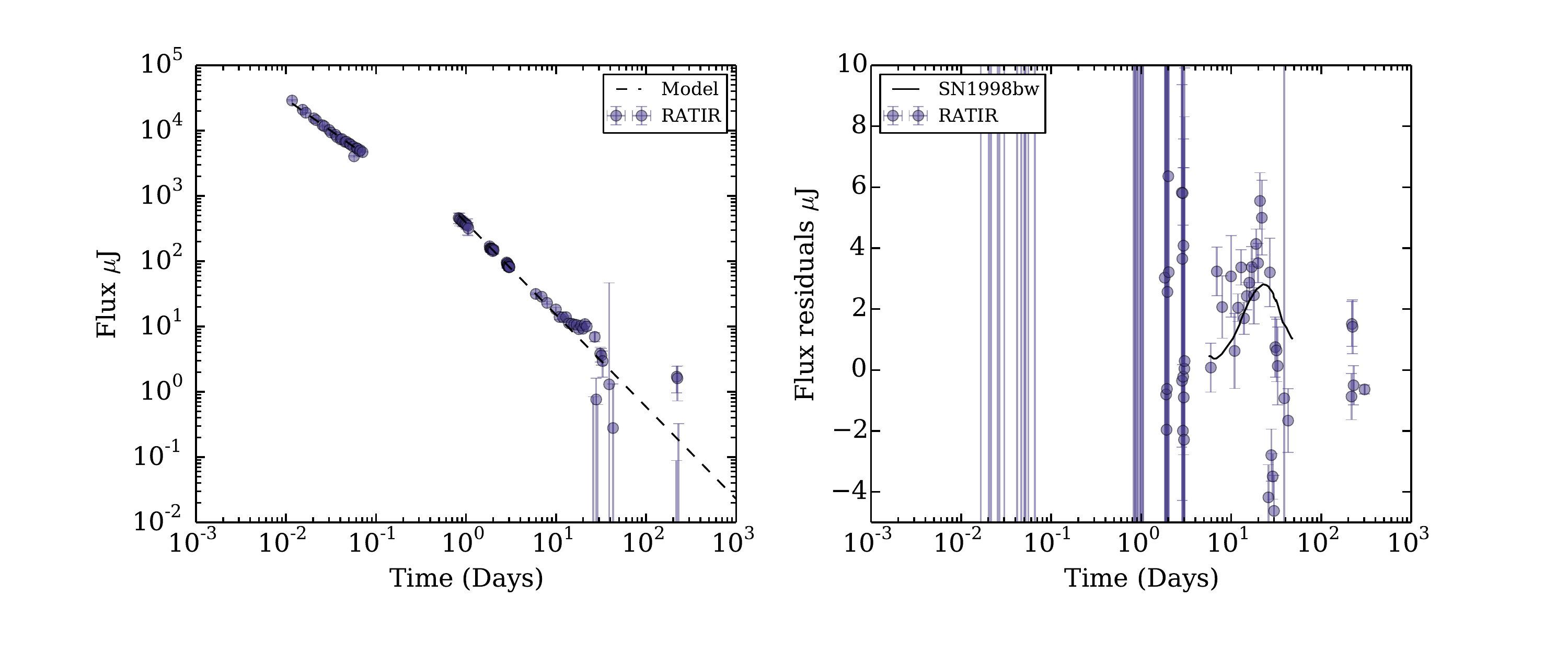}\\[5ex]
 \caption{\emph{Left:} The host-subtracted data (points) and power-law model (dashed line) in $Z$ filter of RATIR (points). \emph{Right:} Flux density residuals in $Z$ (points) and the flux density of SN 1998bw in $R$ shifted to $z = 0.34$ (continuous line).\\}
   \label{fig:sn_Z}
 \end{figure*}

\begin{figure*}
\centering
 \includegraphics[height=0.30\textheight]{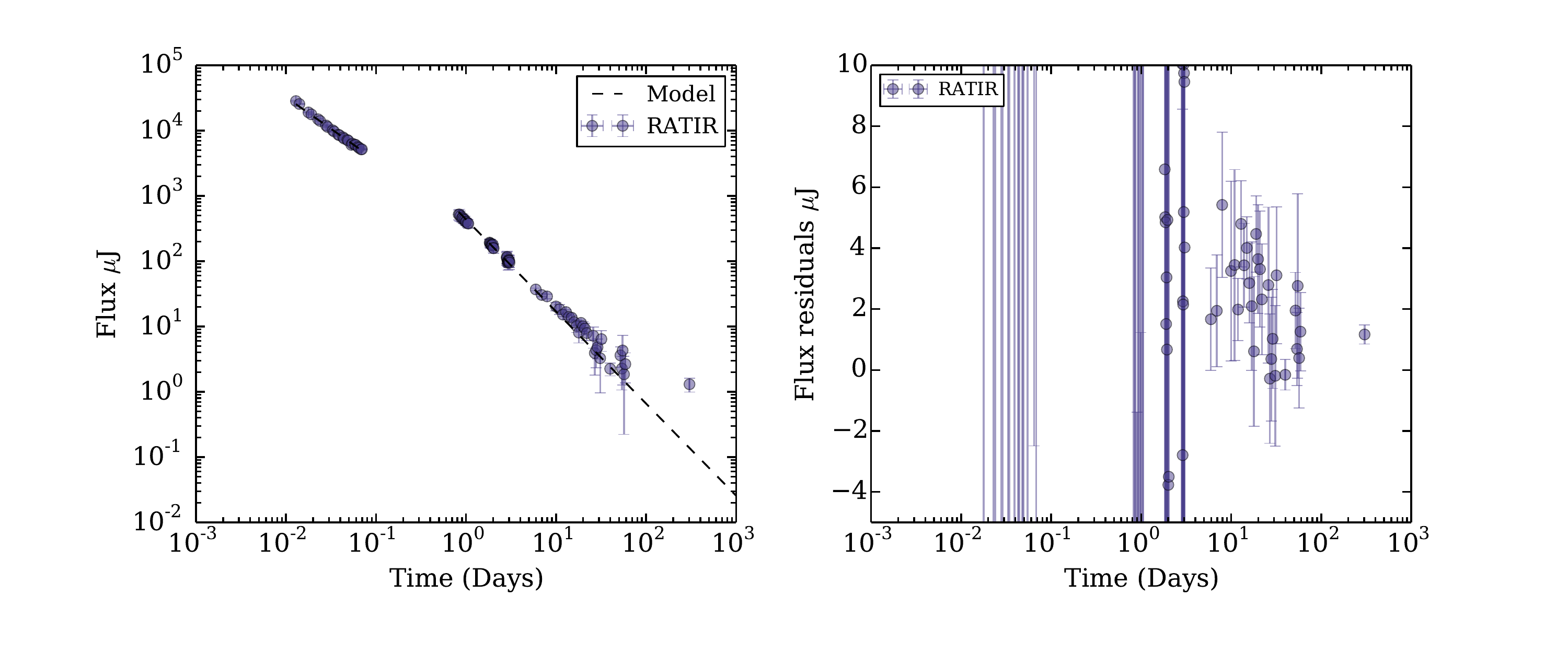}\\[5ex]
 \caption{\emph{Left:} The host-subtracted data (points) power-law model (dashed line) in $Y$ filter of RATIR. \emph{Right:} Flux density residuals in $Y$ (points).\\}
\label{fig:sn_Y}
\end{figure*}

\begin{figure*}
\centering
 \includegraphics[height=0.30\textheight]{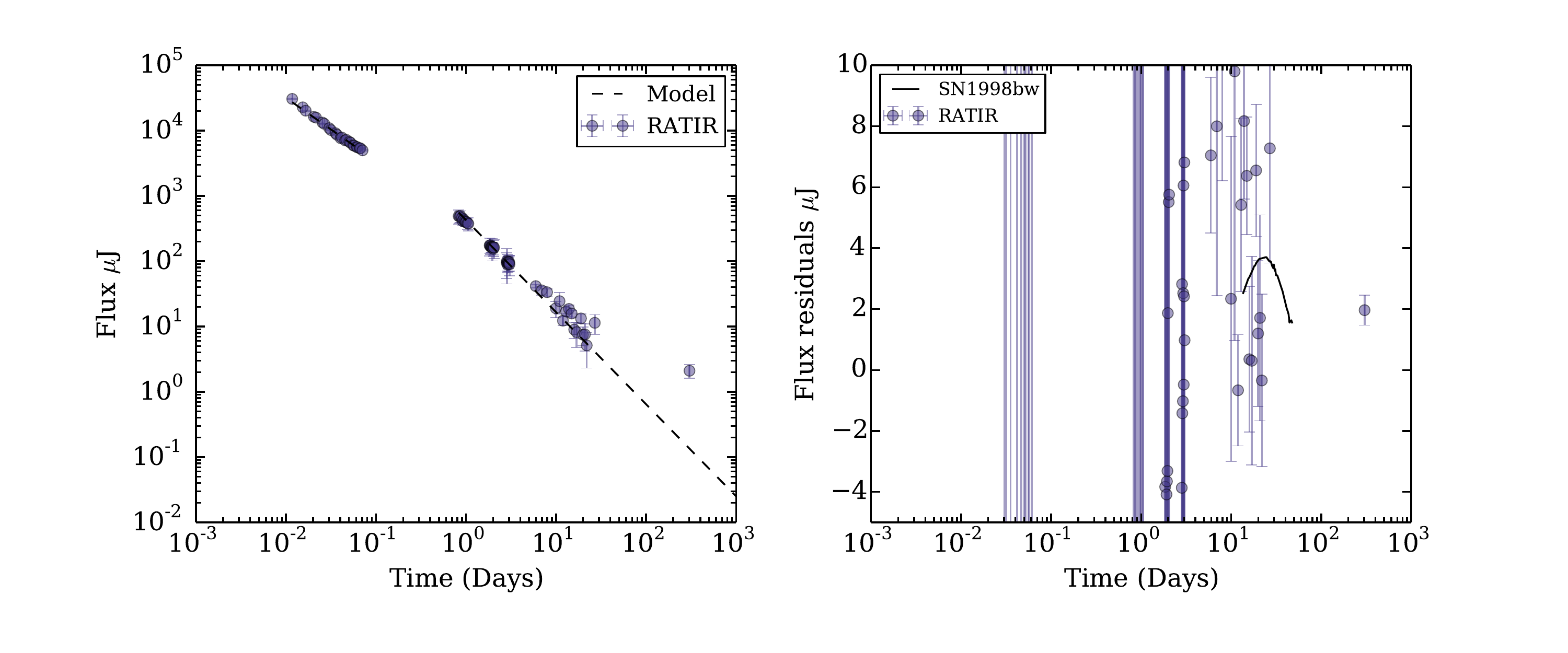}\\[5ex]
 \caption{\emph{Left:} The host-subtracted data (points) and power-law model (dashed line) in $J$ filter of RATIR. \emph{Right:} Flux density residuals in $J$ (points) and the flux density of SN 1998bw in $I$ shifted to $z = 0.34$ (continuous line).\\}
   \label{fig:sn_J}
\end{figure*}

\begin{figure*}
\centering
 \includegraphics[height=0.30\textheight]{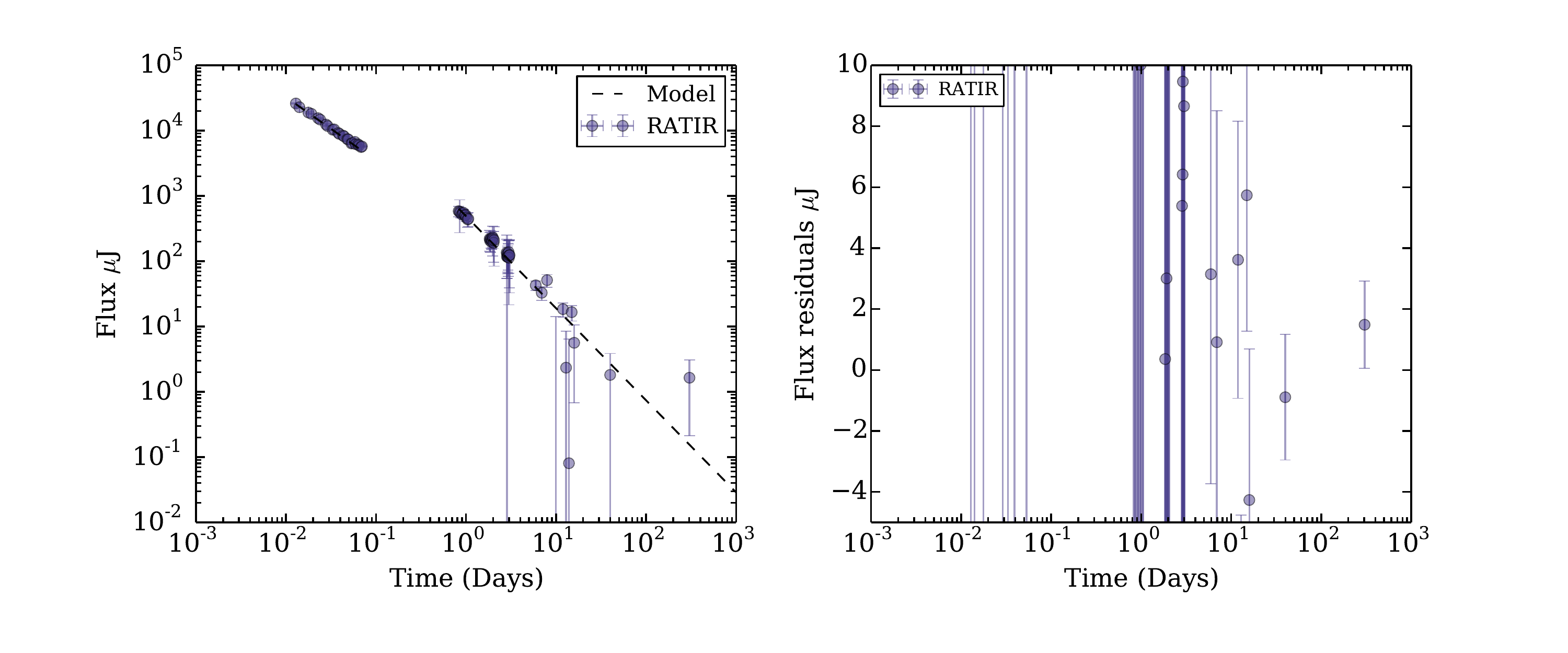}\\[5ex]
 \caption{\emph{Left:} The host-subtracted data (points) and power-law model (dashed line) in $H$ filter of RATIR. \emph{Right:} Flux density residuals in $H$ (points).}
   \label{fig:sn_H}
\end{figure*}

The host-subtracted measurements minus the the best-fit afterglow models (Table \ref{tab:datosres}), henceforth residuals, are show in Figures \ref{fig:sn_g} to \ref{fig:sn_H}. These residuals show a rise and fall from about 7 to about 40 days, confirming the suggestion of \cite{2013GCN..14666...1W}. We propose that this is the photometric signature of SN 2013cq. 

To compare our data to SN 1988bw at a redshift of $z = 0.0085$ \citep{1998IAUC.6896....3W}, we need to account for the effects of redshift on the luminosity distance, observed band, and time dilation. For the luminosity distance, we used a $\lambda$CDM model with a $H_0=67.8$ km/Mpc/s \citep{2014A&A...571A...1P}. The effect of redshift on the filters is shown in Table \ref{tabla:bandas}. In this table, the first and second column give the RATIR filter and its central wavelength $\bar\lambda$ at $z = 0$, the third column gives the central wavelength $\bar\lambda$ in the rest frame of SN 2013cq at $z = 0.34$, and the fourth and fifth give the corresponding Johnson-Cousins filters and their central wavelengths at $z = 0$. Fortuitously, there is a good correspondence between the RATIR $griZYH$ filters at $z = 0.34$ and the Johnson $UBVRIJ$ filters at $z \approx 0$. The time dilation correction is  a factor of $(1+0.34)/(1+0.0085) = 1.33$.

\begin{figure*}
\centering
 \includegraphics[width=0.8\textwidth]{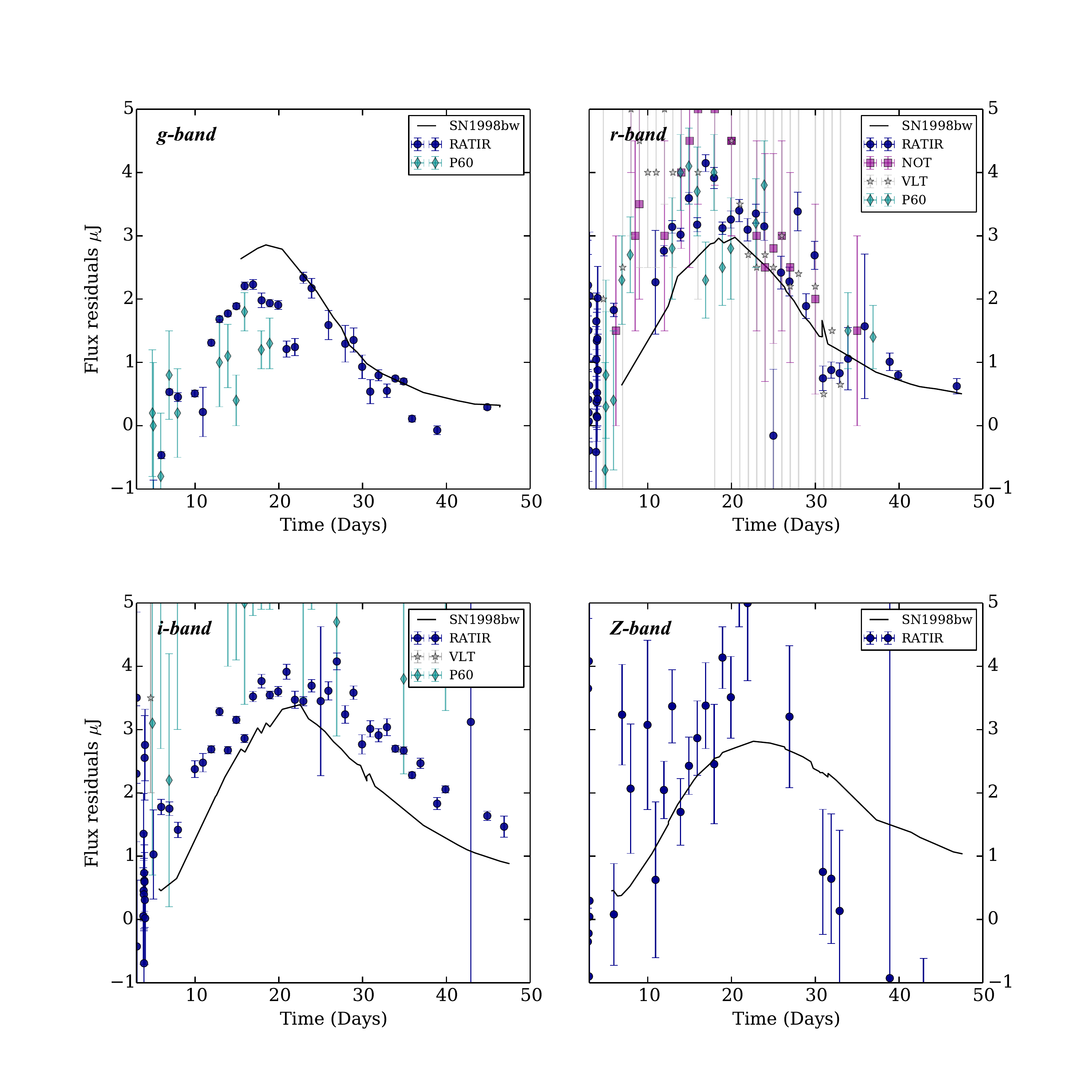}\\[5ex]
 \caption{Light curves of SN 2013cq associated with GRB 130427A in the \emph{griZ} bands. RATIR from this work (points), P60 from \cite{2014ApJ...781...37P} (diamonds), and NOT from \citep{2013ApJ...776...98X} (squares), and VLT from \citep{2014A&A...567A..29M} (stars). The line shows the light curves for SN 1998bw are shifted to $z = 0.34$.}
   \label{fig:sn4}
\end{figure*}

Figure~\ref{fig:sn4} compares our photometry of SN~2013cq with that of SN~1998bw shifted to $z=0.34$ \citep{1998Natur.395..670G,2011AJ....141..163C,2006AA...447..891F}, both bands in rest frame. Qualitatively, the agreement is good, especially in the bluer $griz$ filters although compared with SN~1998bw, SN~2013cq is fainter in $g$-band while is brighter in the $riz$ filters.

 The peak times, calculated by adjusting third-degree polynomials to the residuals (between 7 and 40 days), are 17.66, 17.33 and 22.00 days with $\chi^2/\rm{d.o.f.}$ of 0.62, 0.92 and 0.06 respectively, for the $g$, $r$, and $i$ bands, and are consistent with \cite{2013ApJ...776...98X}. A similar fit to the Z band did not produce a convincing fit (the reduced $\chi^2/\rm{d.o.f.}$ was 5.44) and so we do not have confidence in the peak time for that band.

Moreover, Figure~\ref{fig:sn4} also compares our residuals with the P60 \citep{2014ApJ...781...37P}, NOT \citep{2013ApJ...776...98X} and VLT \citep{2014A&A...567A..29M} photometry. This shows the superior temporal coverage of the RATIR data (we have photometry for every night from nights 5 to 40) and the lower noise.  For example, the errors (associated with the observations) in our residuals, calculated  around the SN peak in \emph{r}-band (between 18 and 26 days after the GRB trigger) are around 0.48~$\mu$Jy, while  
\cite{2014ApJ...781...37P}, \cite{2013ApJ...776...98X}, and \cite{2014A&A...567A..29M} give estimated errors of 1.7, 1.5, and 0.7~$\mu$Jy for their photometry with P60, NOT, and VLT, respectively.

\begin{deluxetable}{lcccc}

\tabletypesize{\scriptsize}
\tablewidth{0pt}
\tablecaption{Correspondence between RATIR filters at $z = 0.34$ and Johnson-Cousins filters at $z \approx 0$\label{tabla:bandas}}
\tablehead{\colhead{Filter}&\colhead{$\bar\lambda$ [nm]}&\colhead{$\bar\lambda$ [nm]} &\colhead{Filter} &\colhead{$\bar\lambda$ [nm]}\\
&\colhead{$z = 0$}&\colhead{$z = 0.34$} &&\colhead{$z = 0$}}
\startdata
$g$&\phantom{0}470&\phantom{0}351&$U$&\phantom{0}360\\
$r$&\phantom{0}618&\phantom{0}461&$B$&\phantom{0}440\\
$i$&\phantom{0}760&\phantom{0}567&$V$&\phantom{0}550\\
$Z$&\phantom{0}878&\phantom{0}655&$R$&\phantom{0}640\\
$Y$&\phantom{}1020&\phantom{0}761&$I$&\phantom{0}759\\
$J$&\phantom{}1250&\phantom{0}932&\nodata&\nodata\\
$H$&\phantom{}1635&\phantom{}1220&$J$&\phantom{}1260
\enddata
\end{deluxetable}

\begin{figure}
\centering
 \includegraphics[width=0.5\textwidth]{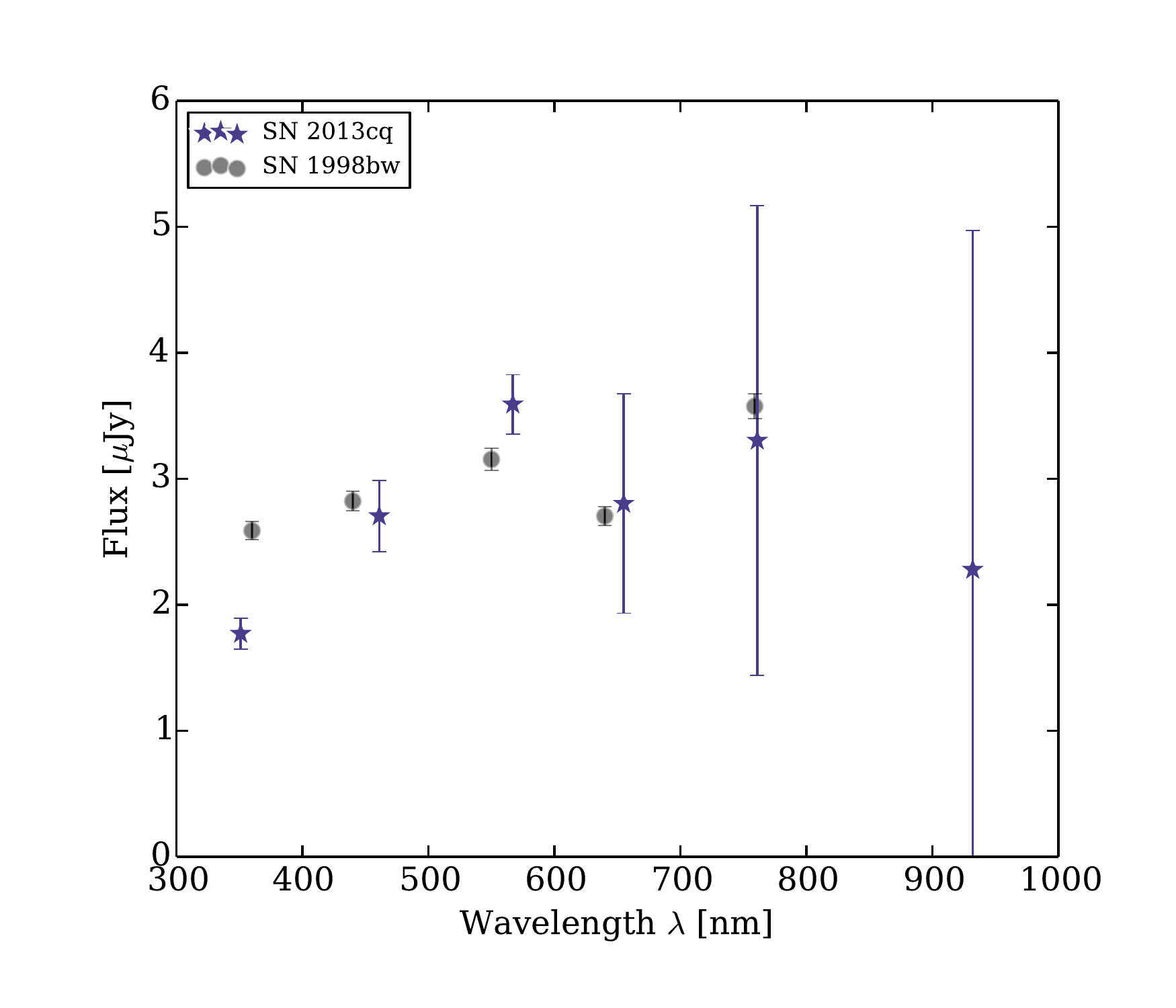}\\[5ex]
 \caption{Broad-band rest-frame SED of SN 2013cq (stars) and SN 1998bw (circles) by averaging their fluxes from days 18 to 26. The wavelength is in the rest frame, but the flux density is observed for SN 2013cq and shifted to $z = 0.34$ for SN 1998bw.}
   \label{fig:sn_spectrum}
\end{figure}

We determined the peak flux density and magnitudes of SN 2013cq and SN 1998bw shifted to $z=0.34$, by averaging the residuals from days 18 to 26. These are given in Table \ref{tabla:mpeak}. The SEDs at the peak flux of both supernovae are shown in Figure \ref{fig:sn_spectrum}. Within our considerable uncertainties, the broadband SED of SN 2013cq is compatible with that of SN 1998bw for 300--800 nm and suggest similarities in the ejected $^{56}$Ni masses and  kinetic  energies between both SNe.
For SN 2013cq the rest-frame $M_r$ magnitude (from our observed  $Z$ magnitude) is  $M_r=-18.43\pm 0.11$ for SN 2013cq, which is close to the measured $M_r=-18.48\pm 0.08$ for SN 1998bw.

\begin{deluxetable*}{lrrrrcrrrrr}
\tabletypesize{\scriptsize}
\tablewidth{0pt}
\tablecaption{Peak flux densities and absolute magnitudes for SN 2013cq and SN 1998bw shifted to $z = 0.34$\label{tabla:mpeak}}
\tablehead{
\colhead{Band}&\multicolumn{2}{c}{{SN 2013cq}}&\multicolumn{2}{c}{{SN 1998bw at $z = 0.34$}}\\
&\colhead{$F_\nu$ [$\mu$Jy]}&\colhead{$M$}&\colhead{$F_\nu$ [$\mu$Jy]}&\colhead{$M$}
}
\startdata
$g$&$+1.77\pm0.12$&$-17.97\pm0.07$&$2.58\pm 0.07$&$-18.38\pm0.03$\\
$r$&$+2.70\pm0.28$&$-18.43\pm0.11$&$2.82\pm 0.08$&$-18.48\pm0.03$\\
$i$&$+3.59\pm0.23$&$-18.74\pm0.07$&$3.15\pm 0.09$&$-18.60\pm0.03$\\
$Z$&$+2.80\pm0.87$&$-18.47\pm0.33$&$2.70\pm 0.07$&$-18.43\pm0.03$\\
$Y$&$+3.30\pm1.86$&$-18.65\pm0.61$&\nodata&\nodata\\
$J$&$+2.28\pm2.69$&$-18.24\pm0.28$&$3.57\pm 0.10$&$-18.74\pm0.03$\\
$H$&\nodata&\nodata&\nodata&\nodata
\enddata
\end{deluxetable*}

\section{Summary}
\label{sec:discussion}

We have presented $griZYJH$ photometry of the afterglow of GRB 130427A with the RATIR instrument from the night of the burst to 3 years later. Comparing our work to previous photometric studies \citep{2014ApJ...781...37P,2013ApJ...776...98X,2014A&A...567A..29M}, we have better temporal sampling, deeper photometry, and we subtract a deep late-epoch image to remove the host galaxy. 

\cite{2014ApJ...781...37P} were better able to study the afterglow over the first couple of days and show the existence of a temporal break at about 0.7 days. Our data, on the other hand, are better suited to looking for deviations from simple power-law model at later times associated with a supernova.

We fit the early afterglow (up to 0.7 days) and late afterglow (after 0.7 days) with power-laws. Prior to fitting for the late afterglow, we subtracted a late image to remove the contamination from the host galaxy. The temporal index of the power-law components changes from 0.97 during the early afterglow to 1.41 during the late afterglow, in agreement with the values and temporal break around 0.7 days determined by \cite{2014Sci...343...358M} and \cite{2014ApJ...781...37P}.

Positive residuals to the fits in $griZ$ between about 7 and 40 days show that we are seeing the photometric signature of SN 2013cq, previously detected spectroscopically by \cite{2013GCN..14646...1D} and \cite{2013ApJ...776...98X} and photometrically by \cite{2014ApJ...781...37P}, \cite{2013ApJ...776...98X}, and \cite{2014A&A...567A..29M}. The absolute magnitude and broadband SED of the supernova are consistent with those of the prototype SN 1998bw and suggest similar progenitors. The peak times agree with the reported by \cite{2013ApJ...776...98X} and is detailed for $griz$ bands. The absolute magnitudes calculated match with SN 1998bw for $riz$ bands. Our better temporal coverage and deeper photometry give us an improved light curve compared to previous work \citep{2013ApJ...776...98X,2014ApJ...781...37P,2014A&A...567A..29M}.     Photometric data obtained three years after the GRB 130427A suggest that the host galaxy is extremely blue compared to local samples.
 
GRB 130427A  is among the handful of events with a confirmed GRB/SN association. In addition it is a high-luminosity event, differing from sub-luminous, very local ones such as GRB980425/SN1998bw. The combination  is  thus unique, and the detailed, homogeneous photometry here presented aims to enlarge the sample for which detailed inferences can be made, eventually in a more statistically significant way when combined with other bursts. Then, the photometry here presented aims to enlarge the sample for which detailed inferences can be made, eventually in a more statistically significant way when combined with other bursts.

\section*{acknowledgments}

We thank the staff of the Observatorio Astron\'omico Nacional on Sierra San Pedro M\'artir. We thank the anonymous referee for a very helpful report. We thank Fabio De Colle for useful comments on earlier drafts. 
RATIR is a collaboration between the University of California, the Universidad Nacional Auton\'oma de M\'exico, NASA Goddard Space Flight Center, and Arizona State University, benefiting from the loan of an H2RG detector and hardware and software support from Teledyne Scientific and Imaging. RATIR, the automation of the Harold L. Johnson Telescope of the Observatorio Astron\'omico Nacional on Sierra San Pedro M\'artir, and the operation of both are funded through NASA grants NNX09AH71G, NNX09AT02G, NNX10AI27G, and NNX12AE66G, CONACyT grants INFR-2009-01-122785 and CB-2008-101958, UNAM PAPIIT grants IG100414 and IA102917, UC MEXUS-CONACyT grant CN 09-283, and the Instituto de Astronom{\'\i}a of the Universidad Nacional Auton\'oma de M\'exico.


\clearpage
\LongTables
\begin{deluxetable*}{rrrrrcrrrrr}
\tablewidth{10pt}
\tablecaption{Photometry of GRB 130427A\label{tab:datos}}
\tablehead{\colhead{$t_0$ (d)}&\colhead{$t_0$ (s)}& \colhead{$t_e$ (s)}&\colhead{$t_f$ (s)}&\colhead{Filter}&\colhead{$m$ (AB)}&\colhead{Flux ($\mu$Jy)}\\
\colhead{}&\colhead{}&\colhead{}&\colhead{}&\colhead{}&\colhead{(aperture)}&\colhead{(PSF)}}

\startdata
\input{datosf.tex}
\enddata
\end{deluxetable*}

\clearpage
\LongTables
\begin{deluxetable*}{rccccccc}
\tablewidth{10pt}
\tablecaption{Residuals between observations and power-law model for RATIR data for early afterglow (before 0.7 days)\label{tab:datosres_early}}
\tablehead{
\colhead{$t_0$ (s)}&
 \multicolumn{5}{c}{Filter}\\[1.2ex]
\colhead{}&
\colhead{$i$ [$\mu$Jy]}&
\colhead{$Z$ [$\mu$Jy]}&
\colhead{$Y$ [$\mu$Jy]}&
\colhead{$J$ [$\mu$Jy]}&
\colhead{$H$ [$\mu$Jy]}}
\startdata
\input{datosres_early.tex}
\enddata
\end{deluxetable*}

\LongTables
\begin{deluxetable*}{rccccccc}
\tablewidth{10pt}
\tablecaption{Residuals between observations and power-law model for RATIR data for late afterglow (after 0.7 days)\label{tab:datosres}}
\tablehead{
\colhead{$t_0$ (d)}&
 \multicolumn{7}{c}{Filter}\\[1.2ex]
\colhead{}&
\colhead{$g$ [$\mu$Jy]}&
\colhead{$r$ [$\mu$Jy]}&
\colhead{$i$ [$\mu$Jy]}&
\colhead{$Z$ [$\mu$Jy]}&
\colhead{$Y$ [$\mu$Jy]}&
\colhead{$J$ [$\mu$Jy]}&
\colhead{$H$ [$\mu$Jy]}}
\startdata
\input{datosres.tex}
\enddata
\end{deluxetable*}

\end{document}

%% file: datosf.tex
0.01&930&80&0.01&$i$&$12.88 \pm 0.02$&\\ 
0.01&1009&67&0.01&$Z$&$12.75 \pm 0.02$&\\ 
0.01&1012&67&0.01&$J$&$12.69 \pm 0.02$&\\ 
0.01&1037&80&0.01&$i$&$12.97 \pm 0.02$&\\ 
0.01&1111&67&0.01&$H$&$12.86 \pm 0.02$&\\ 
0.01&1114&67&0.01&$Y$&$12.77 \pm 0.02$&\\ 
0.01&1141&80&0.01&$i$&$13.12 \pm 0.02$&\\ 
0.01&1216&67&0.01&$Y$&$12.88 \pm 0.02$&\\

%% file: datosres_early.tex
929 &$ 1099.93 \pm 539.14 $&\nodata&\nodata&\nodata&\nodata\\
1009 &\nodata&$ 2805.23 \pm 601.42 $&\nodata&\nodata&\nodata\\
1012 &\nodata&\nodata&\nodata&$ 3354.17 \pm 649.69 $&\nodata\\
1037 &$ 1373.25 \pm 518.59 $&\nodata&\nodata&\nodata&\nodata\\
1111 &\nodata&\nodata&\nodata&\nodata&$ 591.10 \pm 605.02 $\\
1114 &\nodata&\nodata&$ 2951.70 \pm 651.46 $&\nodata&\nodata\\
1140 &$ 314.13 \pm 467.87 $&\nodata&\nodata&\nodata&\nodata\\
1216 &\nodata&\nodata&$ -497.04 \pm 535.12 $&\nodata&$ 2238.44 \pm 567.07 $\\

%% file: datosres.tex
 0.83 &\nodata&$ -23.90 \pm  3.62 $&$ -37.66 \pm  3.51 $&$ -48.43 \pm  9.07 $&$ -50.50 \pm  9.44 $&$ -66.74 \pm 10.87 $&$ -54.80 \pm 10.22 $\\
 0.85 &\nodata&$ -20.88 \pm  4.07 $&$ -38.87 \pm  3.49 $&$ -46.55 \pm  9.88 $&$ -31.91 \pm 10.33 $&$ -56.30 \pm  9.97 $&$ -46.25 \pm 17.28 $\\
 0.87 &\nodata&$ -21.47 \pm  2.53 $&$ -25.19 \pm  3.31 $&$ -41.10 \pm  7.91 $&$ -40.88 \pm  9.58 $&$ -29.30 \pm  9.93 $&$ -41.91 \pm  9.72 $\\
 0.90 &\nodata&$ -15.79 \pm  4.17 $&$ -32.39 \pm  3.86 $&$ -39.28 \pm  8.21 $&$ -56.57 \pm  7.43 $&$ -39.85 \pm  8.11 $&$ -50.19 \pm  7.82 $\\
 0.92 &\nodata&$ -22.70 \pm  2.79 $&$ -28.13 \pm  3.29 $&$ -40.63 \pm  7.88 $&$ -26.98 \pm  7.88 $&$ -69.72 \pm  7.39 $&$ -14.33 \pm  8.76 $\\
 0.94 &\nodata&$ -19.90 \pm  2.81 $&$ -32.23 \pm  3.14 $&$ -35.16 \pm  5.39 $&$ -37.45 \pm  8.05 $&$ -30.91 \pm  8.56 $&$ 14.00 \pm  8.31 $\\
 0.97 &\nodata&$ -14.07 \pm  3.22 $&$ -24.73 \pm  3.38 $&$ -26.06 \pm  6.43 $&$ -28.58 \pm  7.58 $&$ -35.32 \pm  7.58 $&$ -7.64 \pm  7.58 $\\
 0.99 &\nodata&$ -12.95 \pm  3.62 $&$ -25.80 \pm  2.90 $&$ -22.56 \pm  7.51 $&$ -42.19 \pm  6.59 $&$ -38.27 \pm  7.29 $&$ 10.01 \pm  7.96 $\\